\documentclass{aa}
\usepackage{graphics}
\usepackage{txfonts}

\begin{document}

\def\te{T_{e\hskip-0.5pt f\hskip-1.0pt f}}
\def\delt{\scriptstyle \Delta }

\def\eol{\hfil\break}
\def\va{\vskip 5.0 mm} \def\vb{\vskip 2.5 mm} \def\vc{\vskip 1.5 mm}
\def\ha{\hskip 5.0 mm} \def\hb{\hskip 2.5 mm} \def\hc{\hskip 1.5 mm}
\def\vd{\vskip 1.0 mm} \def\hd{\hskip 1.0 mm}

   \title{Theoretical UBVRI colors of iron core white dwarfs }

   \author{J. Madej\inst{1},
           M. Pietrachowicz\inst{1},
           P.C. Joss\inst{2},
           A. Majczyna\inst{3}, 
           A. R\'o\.za\'nska\inst{3},
           M. Nale\.zyty\inst{1} }
   \offprints{M. Pietrachowicz  \\ \email{pietrach@tryton.astrouw.edu.pl} }
   \institute { $^1$ Astronomical Observatory, University of Warsaw,
    Al. Ujazdowskie 4, 00-478 Warszawa, Poland   \\
    $^2$ Department of Physics, Center for Theoretical Physics and Center
       for Space Research, Massachusetts Institute of Technology, \\ 
       \hbox{\ha} Cambridge, MA 02139, U.S.A. \\
    $^3$ Copernicus Astronomical Center, Polish Academy of Sciences,
    Bartycka 18, 00-716 Warsaw, Poland}

   \date{Received ...; accepted ...}

   \titlerunning{Theoretical $UBVRI$ colors of iron core white dwarfs}

   \authorrunning{J. Madej, M. Pietrachowicz, P.C. Joss et al.}

   \abstract{

We explore photometric properties of hypothetical iron core 
white dwarfs and compute their expected colors in $UBVRI$ Johnson
broadband system. Atmospheres of iron core WDs in this paper
consist of pure iron covered by a pure hydrogen layer of an arbitrary 
column mass. LTE model atmospheres and theoretical spectra are calculated 
on the basis of Los Alamos TOPS opacities and the equation of state 
from the OPAL project, suitable for nonideal Fe and H gases. 
We have also computed $UBVRI$ colors of the models and
determined an area on the $B-V$ vs. $U-B$ and $B-V$ vs. $V-I$ planes,
occupied by both pure Fe,
and pure H model atmospheres of WD stars. 
Finally, we search for iron core white dwarf candidates in the available 
literature.


}  \maketitle

%

\section{Introduction}

Chemical composition of white dwarf stars cores has not yet been  
convincingly determined. While most white dwarfs are believed to have 
pure carbon or mixed carbon-oxygen interiors, there are suggestions 
that cores of some WDs are built of heavier elements, and pure 
iron cores represent the extreme case.

An idea that some white dwarfs have cores built of heavier elements 
was put forward by Provencal et al. (1998) and Provencal \& Shipman (1999). 
In particular, they noted that the nearby white dwarf Procyon B has 
relatively normal white dwarf mass of 0.594 $M_\odot$, but has a 
very small radius (cf. also Girard et al. 2000). This would imply that 
Procyon B has a very heavy core, perhaps of iron. They found that radii of 
other WD stars: EG 50 and GD 140 are also located significantly below 
the mass-radius relation predicted for typical carbon interior compositions.
Only recently, Provencal et al. (2002) on the basis of 
precise Hubble spectral measurements recognized that Procyon B
is a normal white dwarf, belonging to the DQZ spectral class. 

Panei et al. (2000) have 
obtained thorough theoretical mass-radius relations for iron-dominated
cores.  They show that EG 50, GD 140 and 40 Eri B satisfy these
relations for iron cores, contrary to mass-radius relations
for carbon and oxygen cores.  Their result convinced us that the existence 
of the iron core white dwarfs is possible.
The above hypothesis is a sufficient reason for us 
to perform and publish this research paper. 

Atmospheres of these stars are probably built from pure iron gas 
covered by hydrogen (and/or helium) layer, since stratified atmospheres 
are common among the known white dwarfs. We expect that most likely
iron and hydrogen are not perfect gases, at least in cooler WDs.

In the following sections we describe model atmosphere calculations
in the local thermodynamical equilibrium (LTE), which are based on a frequently
used equation of state computed by the OPAL Project for nonideal hydrogen 
(Rogers, Swenson, \& Iglesias 1996) and iron gas (Rogers 2002). 
Our calculations use exclusively the Los Alamos TOPS monochromatic 
absorption and scattering opacities, consistent with the
nonideal EOS for both gases (Magee et al. 1995). Model atmosphere 
and spectrum calculations are restricted to atmospheres in radiative 
equilibrium, corresponding to effective temperatures $\te \ge 20000$ K. 

Los Alamos opacity tables were originally designed to study physics
and pulsational instabilities in stellar envelopes by means of Rosseland
mean opacities. Therefore the tables do not have spectral resolution 
adequate to study details of stellar atmospheric spectra. This is the
reason that our paper presents and discusses mostly spectral indices 
$U-B$, $B-V$, $V-R$, and $V-I$ of the broadband Johnson photometric
system. We attempt to find a prescription for differentiating most
common DA white dwarfs with pure hydrogen atmospheres from iron core 
white dwarfs with pure iron or stratified hydrogen--iron atmospheres.

\section{Numerical methods }

\subsection{Model atmosphere computations}

Model atmosphere equations are based on the following equation 
of radiative transfer, suitable for planar geometry 
\begin{equation}
 \mu \, {\partial I_\nu \over {\partial \tau_\nu}} = I_\nu 
 - \epsilon_\nu \, B_\nu -(1-\epsilon_\nu)\, J_\nu \, ,
\label{eq:fer}
\end{equation}
where variables $I_\nu$, $J_\nu$, $B_\nu$ and $\mu$ have their usual 
meaning. The three monochromatic variables are expressed in units 
erg cm$^{-2}$ sec$^{-1}$ Hz$^{-1}$ str$^{-1}$. Dimensionless absorption
is   
$\epsilon_\nu = \kappa_\nu / (\kappa_\nu + \sigma_\nu)$, where  
$\kappa_\nu$ and $\sigma_\nu $ denote LTE true absorption and coherent
scattering coefficients, respectively. 
Absorption coefficient $\kappa_\nu$ has been corrected in our code
for stimulated emission as in LTE. 

Computer code used here has been derived from the code {\sc atm21} adapted
to coherent scattering in stellar atmosphere (exact description in 
Madej 1991, 1994, 1998; Madej \& R\'o\.za\'nska 2000; Stage, Joss, 
\& Madej 2002).

The foregoing papers explain in detail the equations of hydrostatic and
radiative equilibrium, which are imposed on the model atmospheres, as
well as two boundary conditions associated with the equation of
transfer. The equation of transfer and the radiation field was solved
with the method of variable Eddington factors (Mihalas 1978).

Model atmosphere equations were expressed in discrete form on 
frequency grid $\nu_i$, $i=1, \ldots, I$ and standard optical 
depth grid $\tau_d$, $d=1, \ldots, D$. Here, in our models,
the monochromatic optical depth at the fixed wavelength of 1500
{\AA } is used. For each model we  
set $I=749$ and started computations from a very small 
standard optical depth point $\tau_1=10^{-10}$.
We have assumed a value for $D$ equal to $D=150$ for pure hydrogen models,
while for pure iron and stratified H-Fe models $D=155$ was set. 
Both the monochromatic absorption and scattering coefficiens, and hence 
the $\epsilon_\nu$, were determined at each frequency $\nu_i$ and depth 
points $\tau_d$ by direct interpolation of extensive TOPS opacity
tables. The equation of state for nonideal gas, i.e. the values of gas
density for a given temperature and gas pressure, were similarly
determined by interpolation of the OPAL EOS tables either for H or
pure Fe gas.


\subsection{Equation of state and opacity data}

The nonideal equation of state for partially ionized plasma, which
was used in our research, has been extensively described by Rogers (1994),
Rogers, Swenson \& Iglesias (1996), cf. also references therein. 
These authors refer to their
derivation as to the activity expansion of the grand canonical partition
function of plasma, the latter being basic function in statistical
thermodynamics. The method includes quantum mechanical two-body and
many-body interactions, and electron degeneracy along with other effects.
We feel not enough competent to present and discuss details of their method
in the present paper. However, we mention after Rogers et al. (1996) that
their derivation of the EOS for various plasmas basically differs from 
common methods which are based on minimization of free energy in plasma.

This research uses EOS data which is available for a general user.
EOS tables for pure hydrogen plasma were taken from the Lawrence
Livermore National Laboratory www site ({\rm 
http://www-phys.llnl.gov\eol /Research/OPAL/index.html}). With EOS tables
for pure iron gas we were kindly provided by Dr. F.J. Rogers (2002).

Extensive TOPS monochromatic opacity tables have been prepared at the Los Alamos
National Laboratory (Magee et al. 1995). We have collected hydrogen and 
iron data from their web page at 
{\rm http://www.t4.lanl.gov/opacity/tops.html}.
See that www site for more details on the TOPS opacity calculations.


\subsection{Synthetic colors of the Johnson UBVRI system}

Having calculated the models, we acquired their broadband Johnson $UBVRI$
colors using a color synthesis procedure described by Bergeron et al. 
(1995);  see also detailed review of photometric systems by Girardi et al.
(2002).

At first, stellar magnitudes of a theoretical spectrum are computed according
to the standard formula
\begin{equation}
m_{S_\lambda} = - \, 2.5 \, \log  \, {
\int \limits_0^\infty F_\lambda \, S_\lambda \, d\lambda \over {
\int \limits_0^\infty S_\lambda \, d\lambda   }
} \,  + \, m_{S_\lambda^0} \,\, ,
\label{eq:mag}
\end{equation}
where we integrate the product of the flux energy spectrum $F_\lambda$ of a model
atmosphere with the transmission function $S_\lambda$  of the respective 
broadband filter. The constant $m_{S_\lambda^0}$ can assume an arbitrary 
value.

We used the Johnson $UBVRI$ transmission functions defined by Bessell (1990),
and restricted our research to analysis of colors, i.e. differences of the 
respective magnitudes. Then the definition of the respective colors, e.g. $B-V$
color, is as follows:
\begin{equation}
 B-V = -\, 2.5 \, \log \, {
\int\limits_0^\infty F_\lambda S_B (\lambda) \, d\lambda \over  {
\int\limits_0^\infty F_\lambda S_V (\lambda) \, d\lambda  }
}  + C_{B-V}   \, ,
\label{eq:mab}
\end{equation}
cf. Bergeron et al. (1995). Constant $C_{B-V}$ includes all quantities which
are not related to flux $F_\lambda$. 

Note that plane-parallel model atmosphere codes
compute theoretical spectra of radiation $F_\lambda$ which are emitted by unit
surface on the star, this is 1 cm$^2$ in case of the {\sc atm21} code. 
Therefore theoretical flux $F_\lambda$ in Eq.~\ref{eq:mab} is never directly
comparable with e.g. the flux measured on Earth or absolute flux in that
filter, which would be measured at 10 pc distance. 

Eq.~\ref{eq:mab} can be applied to Vega, which is the fundamental 
spectrophotometric standard. In our research each color of Vega is equal to
zero by assumption. Setting $F_\lambda$ to the flux of Vega allowed us 
to determine constants $C_{U-B}$, $C_{B-V}$, $C_{V-R}$ and $C_{V-I}$
for all theoretical colors presented in the following Sections.

For the above calibration purposes we adapted absolute fluxes of Vega
observed by Hayes (1985), for $\lambda > 4500$ {\AA }. His paper gives
values of the Vega flux in equidistant intervals of 25 {\AA } and
therefore gives poor description of the flux in the near ultraviolet, where
prominent higher Balmer lines merge and the wavelength depence of Vega
flux is very complex. Therefore for $\lambda < 4500$ {\AA } we have taken
theoretical fluxes from Castelli \& Kurucz (1994). They have computed
{\sc atlas12} model atmosphere and spectrum of Vega with a high spectral
resolution, normalised in the same way as Hayes' (1985) observations.

We have taken this approach to reference spectra of Vega from Bergeron
et al. (1995), in order to make our color indices for hydrogen and iron
atmospheres fully compatible with their tables of Johnson photometry colors
computed for DA and DB white dwarfs.

\section{UBVRI colors of pure H atmospheres }

In this section we construct and discuss two different series of
pure hydrogen model atmospheres of white dwarfs. This is done
in order to compare and measure the discrepancies between each other
and assess the uncertainty of colors of atmospheres computed with the
following two basically different numerical methods. By doing this,
we also test the compatibility of both OPAL and 
TOPS data with the existing model atmosphere codes. If effect, we draw
a conclusion that our pure Fe model atmospheres, based on those data, 
are consistent with pure H atmospheres. 

For calculating models of hot WD atmospheres consisting of 
pure hydrogen, the following approaches were undertaken:
\eol 
(1) We assumed that plasma of hot hydrogen atmosphere is 
described by the EOS of ideal gas.  Under this 
assumption, thorough models were calculated in the 
NLTE regime.  Models and spectra computations were performed 
using {\sc tlusty195} and {\sc synspec42} codes (Hubeny 1988; Hubeny \& Lanz
1992, 1995).  
Three spectral series of hydrogen (Lyman, Balmer, and Paschen) were 
considered as we computed outgoing flux spectra and color indices.
Since Paschen lines are not present in the standard versions of both
codes, we have adopted to the codes the opacities of thermally \& 
pressure-broadened Paschen lines from Lemke (1997).
\eol
(2) Alternatively, we assumed more factual, nonideal EOS for gas 
and derived a number of models in LTE.  Tables for nonideal 
EOS were taken from the OPAL project, while the Los Alamos TOPS 
monochromatic opacity tables were used.

\begin{table}
\caption{Broadband colors in stellar magnitudes for pure hydrogen 
atmospheres. Models are computed also with nonideal EOS and Los Alamos 
(TOPS) opacities. Temperatures $\te$ are given in thousands K and
$\log g $ are in cgs units.}
\label{tab:col1}
\renewcommand{\arraystretch}{1.0}
\linespread{0.5}
\begin{tabular}{|ccccrcc|}
    \hline
$Z$ &  $\te$ & $\log g$ &  $U-B$ & $B-V$ & $V-R$ & $V-I$ \\
\hline \hline 
1 &   20 &  7 &  -0.895  &  -0.057  &   -0.098 &  -0.206 \\
1 &   20 &  8 &  -0.965  &  -0.012  &   -0.103 &  -0.202 \\
1 &   20 &  9 &  -1.017  &   0.014  &   -0.109 &  -0.199 \\
1 &   20 & 10 &  -1.076  &   0.026  &   -0.112 &  -0.180 \\[5pt]
1 &   30 &  7 &  -1.121  &  -0.193  &   -0.135 &  -0.290 \\
1 &   30 &  8 &  -1.157  &  -0.171  &   -0.139 &  -0.288 \\
1 &   30 &  9 &  -1.186  &  -0.163  &   -0.142 &  -0.288 \\
1 &   30 & 10 &  -1.211  &  -0.168  &   -0.145 &  -0.275 \\[5pt]
1 &   50 &  7 &  -1.238  &  -0.284  &   -0.146 &  -0.325 \\
1 &   50 &  8 &  -1.244  &  -0.277  &   -0.147 &  -0.325 \\
1 &   50 &  9 &  -1.251  &  -0.275  &   -0.148 &  -0.323 \\ 
1 &   50 & 10 &  -1.256  &  -0.275  &   -0.149 &  -0.320 \\[5pt]
1 &   70 &  7 &  -1.257  &  -0.313  &   -0.145 &  -0.338 \\
1 &   70 &  8 &  -1.262  &  -0.308  &   -0.150 &  -0.339 \\
1 &   70 &  9 &  -1.268  &  -0.304  &   -0.150 &  -0.336 \\
1 &   70 & 10 &  -1.271  &  -0.304  &   -0.152 &  -0.336 \\[5pt]
1 &   100 & 7 &  -1.279  &  -0.329  &   -0.142 &  -0.348 \\
1 &   100 & 8 &  -1.280  &  -0.327  &   -0.145 &  -0.350 \\
1 &   100 & 9 &  -1.282  &  -0.324  &   -0.146 &  -0.349 \\
1 &   100 &10 &  -1.282  &  -0.324  &   -0.150 &  -0.349 \\
\hline
\end{tabular}
\end{table}

\begin{table}
\caption{Broadband colors in stellar magnitudes for pure 
hydrogen NLTE atmospheres (results of {\sc tlusty 195}). Units
are the same as in Table 1.  }
\label{tab:col2}
\renewcommand{\arraystretch}{1.0}
\begin{tabular}{ccccrc}
    \hline
$Z$ &  $\te$ & $\log g$ &  $U-B$ & $B-V$ & $V-I$ \\
\hline \hline 
1 &   20 &  7 &  -0.812  &  -0.091 & -0.209 \\
1 &   20 &  8 &  -0.866  &  -0.035 & -0.210 \\
1 &   20 &  9 &  -0.927  &   0.029 & -0.203 \\[5pt]
1 &   50 &  7 &  -1.217  &  -0.291 & -0.330 \\
1 &   50 &  8 &  -1.224  &  -0.278 & -0.331 \\
1 &   50 &  9 &  -1.229  &  -0.258 & -0.327 \\[5pt]
1 &   100 & 7 &  -1.270  &  -0.328 & -0.350 \\
1 &   100 & 8 &  -1.271  &  -0.322 & -0.352 \\
1 &   100 & 9 &  -1.269  &  -0.309 & -0.348 \\
\hline
\end{tabular}
\end{table}

In the present work we are concerned about isolated hot WD 
atmospheres with effective temperatures $\te$ in the 
range of 20000 to 100000 K and surface gravities ranging from
$\log g = 7.0$ to $10.0$ (cgs units). White dwarfs of as high 
surface gravity as $\log g = 10.0$ are not yet known, however, one can
expect that iron core white dwarfs of a given mass have a particularly 
small radius and therefore high surface gravity. Consequently, we 
appended high gravity hydrogen model atmospheres to obtain grid
consistent with pure iron high gravity atmospheres. 
The radiative and hydrostatic equilibria are implied in all models.  

Grids of color indices for both NLTE ({\sc tlusty195}) and LTE (Los Alamos) 
series of models were plotted on two-index plane, $U-B$ vs. $B-V$,  
and are displayed on Figure 1.  Colors of the TOPS-based LTE 
models are in a reasonable agreement 
with the corresponding colors of the NLTE models, especially for highest
effective temperatures $\te$.  The largest difference in the 
$U-B$ index is equal to almost 0.1 mag (for $\te = 20000$ K and $\log g = 8$);  
in the $B-V$ it is mere 0.034 mag (for $\te = 20000$ K and $\log g =
7$). The differences in $V-I$ color are smaller still.

We conclude that the models, based on TOPS and OPAL tables,
are in satisfactory agreement with the NLTE {\sc tlusty195} models,
as long as pure hydrogen models are concerned.  Therefore we 
believe that pure iron or stratified hydrogen-iron models with nonideal
EOS would be consistent with NLTE ideal gas models at the high
$\te$ limit.  

It is interesting to note that colors of pure H model atmospheres
computed either with the EOS of ideal gas or with the OPAL nonideal EOS
exhibit systhematic differences for lower effective temperatures $\te$ 
of our grid. Only at the highest $\te$ both series of models converge
on the $U-B$ vs. $B-V$ plane. Such systhematic differences are not present
in the $B-V$ vs. $V-I$ diagram and hence are not displayed here.


   \begin{figure*}
    \resizebox{12cm}{!}{\includegraphics{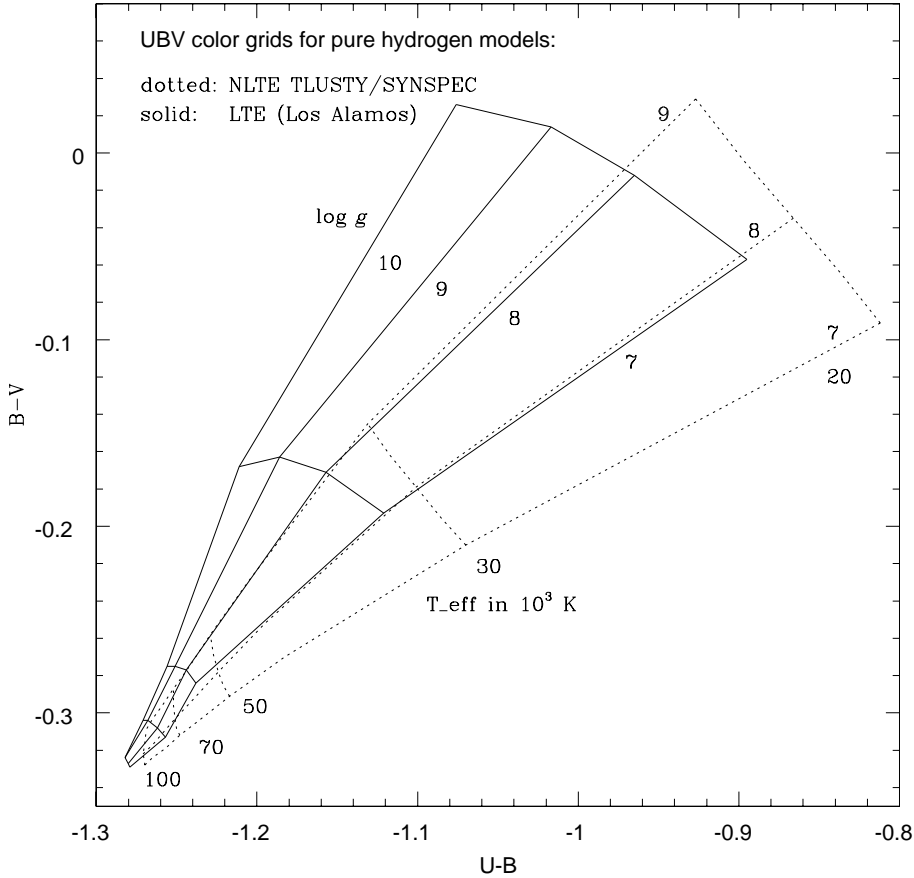}}
    \hfill
    \parbox[b]{55mm}{   
    \caption{Comparison of theoretical colors, $B-V$ vs. $U-B$, for two
    grids of pure hydrogen, hot DA atmospheres. Models are computed either
    with the {\sc tlusty195} and {\sc synspec42} NLTE codes, or with our
    {\sc atm21} code based on the nonideal EOS and Los Alamos opacities. 
    Consistency of both families of broadband $B-V$
    and $U-B$ indices is not perfect but is satisfactory.} }
   \end{figure*}

Furthermore, our synthetic NLTE colors of pure hydrogen white dwarfs are
nearly identical with both our LTE colors of these stars (computed with
other option to the {\sc tlusty195} code and still using ideal EOS) 
and the LTE colors obtained 
by Bergeron et al. (1995) in their intensive study on ages, luminosities,
and other photometric parameters of DA white dwarfs.

\subsection{ Discussion of discrepancies }

The above-mentioned systhematic differences could be caused by the effects 
of deviations from the ideal EOS for hydrogen gas, since they vanish at
the highest $\te$. In order to verify such a hypothesis, we have
chosen several temperature $T$ and gas pressure $P_g$ points across our
sample pure H model atmosphere, $\te=20000$ K and $\log g =8.0$, computed
with the nonideal EOS and TOPS opacity data, see Table~\ref{tab:model}.

We compared two different densities corresponding to the same pair of
$T$, $P_g$: density $\rho_{ROG}$, obtained by direct interpolation of
the OPAL tables, and $\rho_{ID}$ corresponding to an ideal H gas, obtained
from the relevant subroutines of the {\sc atm21} code. One can see that
the difference between both densities reach 0.3 \% at the bottom of the
model, and it is much smaller in upper layers. Therefore we conclude
that the EOS of an ideal gas is well fulfilled in all our pure H atmospheres,
including also models of higher $\te$, of course. This firmly suggests 
that the impact of nonideal gas effects on the color discrepancies between
our H models and these obtained with {\sc tlusty195} is negligible.

The referee pointed out that such differences could be due to NLTE effects
considered by the code {\sc tlusty195}, or other reasons.  Earlier studies
on this subject have already predicted the nonideal EOS's and NLTE 
effects in hydrogen atmospheres of white dwarfs. Wesemael et al. (1980)
in their extensive paper on WD model atmospheres and spectra, have also
studied the influence of hydrogen line blanketing and NLTE effects on
both $U-B$ and $B-V$ color indices. They found, for a pure H model
atmosphere of $\te = 50000$ K and $\log g =8.0$, that both ${\delt} (U-B)$ and
${\delt} (B-V)$ line-blanketing corrections are of the order 0.001 mag (at most).

Therefore NLTE corrections to UBV are just negligible in case of pure H
white dwarfs, for relatively hot objects of $\te \ge 20000$ K. This 
conclusion is also supported by our own computations, which were not 
detailed in this paper.

The influence of blanketing by Balmer and Lyman lines is much more 
important. Wesemael et al. (1980) have determined that both ${\delt} (U-B)$
and ${\delt} (B-V)$ corrections, LTE minus NLTE, are equal to $-$0.017 mag and
+0.043 mag, respectively, for a model of $\te = 50000$ K and $\log g =8.0$.
Both corrections should increase in their absolute value with decreasing
$\te$ (see p. 300 of that paper). The quantity of both corrections, when
extrapolated down to the $\te=20000$ K, seems comparable with both differences
${\delt} (U-B)$ and ${\delt} (B-V)$, cf. Fig. 1. 

Note that both series of pure H model atmospheres displayed in Fig. 1
take into account the blanketing by Balmer and Lyman lines.  However, one can
expect that perhaps TOPS hydrogen opacity tables do not include most 
sophisticated theory of H line broadening, thus causing ${\delt} (U-B)$
and ${\delt} (B-V)$ errors seen in Fig. 1. The difference strongly decreases
in $V-I$ color, since in both $V$ and $I$ filters there are only single
Balmer lines, and their broadening gets unimportant for the wide-band 
photometry.

At this point we turn the reader's attention to the paper by Hubeny, Hummer 
\& Lanz (1994), who discuss theory of NLTE model atmospheres with line
blanketing also near the series limit (the occupation probability formalism).
The latter implies dissolving of merging lines, which is the effect of 
nonideal gas in a stellar atmosphere (see also Lanz \& Hubeny, who applied
theory of dissolving levels in hydrogen dominated DA atmospheres with 
carbon or iron in trace abundances). Such a nonideal gas effect has been
included also in our {\sc tlusty195} hydrogen atmosphere calculations, 
see Table 2 and Fig. 1. 

Recent paper by Rohrmann et al. (2002) presents theory and derivation of
the nonideal EOS for H and He gas, including their molecules. 
Rohrmann et al. (2002) concentrate on the theory of He core white dwarfs,
which are much colder than WDs discussed in this paper.

\section{Pure iron and H-Fe stratified atmospheres }

We have computed and present here extensive set of pure iron 
model atmospheres and synthetic $UBVRI$ broadband colors for models of
$20000 \le \te \le 100000$ K and $\log g=7,8,9,$ and 10 (cgs). 
We have extended our
computations to an extremely large $\log g$. In fact we are not aware of 
the existence of any white dwarf star with $\log g \ge 10$ 
but assume that iron cores of a given mass $M$ would have much
smaller radius $R$, and therefore $\log g$, than common carbon-oxygen cores.
Pure H model atmospheres discussed in the previous section were 
also extended to such a large surface gravity due to the same reason.

Pure Fe model atmospheres were computed using nonideal EOS by Rogers (2002)
and Los Alamos TOPS monochromatic opacities. Iron models were computed also
on the grid of 155 discrete standard optical depth points, starting from
the extremely high layer $\tau_1=10^{-10}$ and with 901 discrete 
frequencies. Computations were very difficult and sometimes unstable due to
complicated run of monochromatic opacities in Los Alamos tables and due to
high gradients of opacities (with respect both to temperature and
gas density) at higher densities of this exotic chemical composition. 
Moreover, pure iron model of $\te=100000$ K and $\log g =7$ was not in 
hydrostatic equilibrium, since the model exceeded the Eddington limit 
in its deeper layers. In other words, the acceleration of gas, exerted by
the integrated
radiative flux and directed upwards, exceeded gravitational acceleration
$g$ in those layers.

\subsection{Physical conditions in Fe atmospheres }

Physical properties of pure Fe white dwarf atmospheres do not differ
substantially from pure H atmospheres, at least for relatively high
effective temperatures $\te \ge 20000$ K which were investigated in this
paper. Table~\ref{tab:model} presents concisely values of few most important
thermodynamical variables across the sample pure Fe model atmosphere
with $\te=20000$ K and $\log g =8.0$, organized in the same way as it was
described in Section 3.1.

After each temperature iteration (and also after the final one), gas
presures $P_g$ in the model are determined by integrating the equation
of hydrostatic equilibrium
\begin{equation}
{{d P_g }\over {d\tau_{std}}} = {g \over {(\kappa_\nu+\sigma_\nu)_{std} }}
  - {dP_{rad} \over {d\tau_{std}}} \, ,
\end{equation}
where all the monochromatic quantities were computed at the standard 
wavelength, $\lambda_{std}=1500$ {\AA }. One can note that gas pressure
$P_g(\tau_{std})$ is essentially determined by run of the standard opacities,
$(\kappa_\nu+\sigma_\nu)_{std} $.

Column No. 6 in Table~\ref{tab:model} shows that TOPS opacities (for 1
gram of gas) are only $\approx 4$ times larger for iron than for hydrogen.
This implies that resulting gas pressures $P_g$ on a fixed $\tau_{std}$
are few times smaller in pure Fe sample atmosphere than in pure H
atmosphere of the same $\te=20000$ K. Iron and hydrogen atmospheres do not
differ substantially from each other in the run of physical variables 
$T$, $P_g$, and the density $\rho$.

On the contrary to the pure H model, columns 4 and 5 of Table~\ref{tab:model}
show that the realistic nonideal Fe gas density $\rho_{ROG}$ (Rogers 2002)
differs
from the density $\rho_{ID}$, computed by our {\sc atm21} code according
to the ideal gas EOS. The difference of both densities reaches one order of
magnitude in that model, which clearly demonstrates that the ideal EOS of iron
would be of no use in WDs of $\te=20000$ K. 
We have checked that iron gas of density $\rho = 10^{-5} $ (cgs) behaves 
like an ideal gas in temperatures $T$ as high as $10^6$ K or higher.

Column 7 in Table~\ref{tab:model} presents geometrical depths of various standard
optical depth levels $\tau_{std}$, assuming that the zero (reference)
level of geometrical height is fixed at the upper boundary of a model
atmosphere. The depth of a fixed level $\tau_{std}$ is given by
\begin{equation}
HGT = - \int\limits_0^{\tau_{std}} {d\tau \over {\rho\,
    (\kappa_\nu+\sigma_\nu)_{std} }} \, .
\end{equation}
The total thickness of a sample pure H model atmosphere of $\te=20000$
K and $\log g =8.0$ equals to 7 kilometers. The corresponding pure Fe 
atmosphere is thinner,
obviously due to a higher molecular weight of an iron gas.

\begin{table*}
\caption{Model atmospheres of $\te=20000$ K, and $\log g=8.0$ (cgs units),
   with pure hydrogen or pure iron composition. }
\label{tab:model}
\renewcommand{\arraystretch}{1.0}
\begin{tabular}{ccccccc}
\hline \hline  \\[-3mm]
\noindent{ H model }\hfil  \\[2pt]
$\tau_{std}$& $T$    &   $P_g$  &$\rho_{ROG}$&$\rho_{ID}$&$OPSTD$&$HGT -$ cm\\[1mm]
1.0000E-10&1.1787E+04&1.7938E-02&9.2253E-15&9.2241E-15&5.5747E-01&0.0000E+00\\
1.0000E-06&1.1787E+04&9.6360E+01&4.9551E-11&4.9611E-11&1.5151E+00&1.6705E+05\\
1.0000E-03&1.3906E+04&6.3030E+03&2.7472E-09&2.7560E-09&2.7532E+01&2.5270E+05\\
1.0000E-01&1.7834E+04&1.0314E+05&3.5055E-08&3.5228E-08&1.4934E+02&3.2323E+05\\
1.0000E+00&2.5032E+04&6.0435E+05&1.4629E-07&1.4709E-07&2.4124E+02&3.8450E+05\\
1.0000E+02&6.0790E+04&2.3710E+07&2.3641E-06&2.3726E-06&6.4096E+02&6.3574E+05\\
2.6102E+02&6.7564E+04&4.5851E+07&4.1132E-06&4.1262E-06&8.0050E+02&7.0605E+05\\[2mm]
\noindent{ Fe model  }  \\[2pt]
$\tau_{std}$& $T$    &   $P_g$  &$\rho_{ROG}$&$\rho_{ID}$&$OPSTD$&$HGT -$ cm\\[1mm]
1.0000E-10&1.1411E+04&1.6666E-01&3.6337E-13&3.2686E-12&5.9939E-02&0.0000E+00\\
1.0000E-06&1.2684E+04&1.6718E+02&3.2779E-10&2.9763E-09&1.0046E+00&3.2555E+04\\
1.0000E-03&1.4567E+04&9.1482E+03&1.5621E-08&1.4709E-07&3.8319E+01&5.4919E+04\\
1.0000E-01&1.7476E+04&6.0067E+04&8.5492E-08&7.8795E-07&2.8835E+02&6.7055E+04\\
1.0000E+00&2.2247E+04&2.2637E+05&2.5312E-07&2.2825E-06&5.7911E+02&7.7449E+04\\
1.0000E+02&5.5742E+04&9.6135E+06&4.2901E-06&2.4393E-05&1.2986E+03&1.3145E+05\\
2.6102E+02&6.3817E+04&1.7025E+07&6.6363E-06&3.6046E-05&3.4230E+03&1.4520E+05\\[2pt]
\hline
\end{tabular}
\end{table*}

\subsection{Spectra of Fe atmospheres}

Sample spectra of both pure iron and pure hydrogen model atmospheres are
plotted in Figs 2-7. They are arranged in order of increasing effective 
temperatures $\te$ to demonstrate quantitatively evolution of both Fe and
H spectra with increasing $\te$ in the wide spectral region where the
most of radiative energy is emitted. Effective temperatures $\te$ chosen 
for this presentation range from 20000 K up to 100000 K and $\log g = 8.0$
or 9.0 in hotter models.

Both series of model atmospheres were computed with the nonideal EOS from
the OPAL Project to ensure consistency. 
One can see that pure Fe spectrum is entirely different from the H spectrum,
both in visual (Fig. 3) and in far and extreme UV (Figs. 2 and 4-7). 
The Los Alamos TOPS monochromatic opacities suitable to nonideal EOS allowed
us to reproduce strongly broadened hydrogen Lyman and Balmer lines (Fig. 3).
Appearance of theoretical optical and near UV spectrum of the iron
atmospheres is completely different and also exhibits strongly broadened iron
features, but their identification is beyond the scope of this paper.

Fig. 3 shows that the sample pure Fe atmosphere of $\te = 20000$ K is
much brighter than pure H atmosphere in visual and also in far UV region
beyond the Lyman edge (wavelengths $\lambda < 912$ {\AA }). In near UV,
for $\lambda > 912$ {\AA }, pure Fe atmosphere is fainter that its
pure H (i.e. the DA type) counterpart. Figs. 4-7 demonstrate that Fe
atmospheres are very bright for $\lambda < 912$ {\AA } also in higher
$\te$. Only in the extreme UV, for $\lambda < 300 - 400 $ {\AA },
Fe atmospheres get extremely faint apparently due to increasing iron
monochromatic absorption.


The latter effect, i.e. an increase of iron absorption in EUV and X-rays,
is clearly seen in TOPS tables at any fixed temperature $T$
and gas density $\rho$. In general, iron gas is much more opaque than
hydrogen at any wavelength. This implies also that our pure Fe model 
atmospheres are geometrically thinner, as compared with pure H atmospheres
of the same $\te$ and $\log g$, see also Table~\ref{tab:model}. 


\subsection{Broadband UBVRI colors}

Theoretical broadband $UBVRI$ colors of our pure iron models are 
displayed in Table 3. Comparison of Tables 1 and 3 shows that $U-B$ 
and $B-V$ colors of iron models are significantly lower than the same 
color indices of hydrogen model atmospheres. Therefore, a pure iron 
atmosphere of a given $\te$ and $\log g$ appears as if it was much 
hotter than pure H atmosphere with the same parameters
(an iron WD appears more blue to an observer).

\begin{table}
\caption{Broadband colors in stellar magnitudes for pure iron atmospheres.
Models are computed with nonideal EOS and Los Alamos (TOPS) opacities.
Units of $\te$ and $\log g$ are the same as in Table 1. }
\label{tab:col3}
\renewcommand{\arraystretch}{1.0}
\begin{tabular}{|cccccrc|}
    \hline
$Z$ &  $\te$ & $\log g$ &  $U-B$ & $B-V$ & $V-R$ & $V-I$ \\
\hline \hline 
26 &  20 &  7 & -1.330 &   -0.284   &  0.097 &  -0.063 \\
26 &  20 &  8 & -1.188 &   -0.191   &  0.011 &  -0.133 \\
26 &  20 &  9 & -1.110 &   -0.111   & -0.047 &  -0.185 \\
26 &  20 & 10 & -1.094 &   -0.088   & -0.040 &  -0.191 \\[5pt]
26 &  30 &  7 & -1.432 &   -0.225   & -0.083 &  -0.287 \\
26 &  30 &  8 & -1.432 &   -0.214   & -0.108 &  -0.278 \\
26 &  30 &  9 & -1.418 &   -0.214   & -0.137 &  -0.284 \\
26 &  30 & 10 & -1.377 &   -0.231   & -0.139 &  -0.279 \\[5pt]
26 &  50 &  7 & -1.360 &   -0.363   & -0.177 &  -0.432 \\ 
26 &  50 &  8 & -1.376 &   -0.313   & -0.194 &  -0.467 \\
26 &  50 &  9 & -1.431 &   -0.256   & -0.202 &  -0.459 \\
26 &  50 & 10 & -1.440 &   -0.235   & -0.188 &  -0.418 \\[5pt]
26 &  70 &  7 & -1.322 &   -0.387   & -0.174 &  -0.420 \\
26 &  70 &  8 & -1.311 &   -0.396   & -0.188 &  -0.443 \\
26 &  70 &  9 & -1.303 &   -0.379   & -0.193 &  -0.459 \\
26 &  70 & 10 & -1.313 &   -0.340   & -0.197 &  -0.467 \\[5pt]
26 &  100&  7 & -1.323 &   -0.376   & -0.168 &  -0.423 \\
26 &  100&  8 & -1.315 &   -0.358   & -0.133 &  -0.321 \\
26 &  100&  9 & -1.327 &   -0.419   & -0.187 &  -0.442 \\
26 &  100& 10 & -1.321 &   -0.425   & -0.195 &  -0.461 \\
\hline
\end{tabular}
\end{table}

\begin{table}
\caption{Broadband colors in stellar magnitudes for stratified
hydrogen/iron models with nonideal EOS and Los Alamos (TOPS) opacities.
Units of $\te$ and $\log g$ are the same as in Table 1. 
Transition depth is located at $\tau=10^{-3}$.  }
\label{tab:col4}
\renewcommand{\arraystretch}{1.0}
\begin{tabular}{|cccccrc|}
    \hline
$Comp.$ &  $\te$ & $\log g$ &  $U-B$ & $B-V$ & $V-R$ & $V-I$ \\
\hline \hline 
H/Fe &  20 &  7 & -1.300 &   -0.262   &  0.080 &  -0.057 \\
H/Fe &  20 &  8 & -1.174 &   -0.180   & -0.024 &  -0.141 \\
H/Fe &  20 &  9 & -1.095 &   -0.097   & -0.080 &  -0.192 \\
H/Fe &  20 & 10 & -1.098 &   -0.091   & -0.071 &  -0.197 \\[5pt]
H/Fe &  30 &  7 & -1.433 &   -0.216   & -0.088 &  -0.287 \\
H/Fe &  30 &  8 & -1.435 &   -0.207   & -0.116 &  -0.282 \\
H/Fe &  30 &  9 & -1.417 &   -0.209   & -0.143 &  -0.288 \\
H/Fe &  30 & 10 & -1.378 &   -0.229   & -0.147 &  -0.285 \\[5pt]
H/Fe &  50 &  7 & -1.363 &   -0.357   & -0.177 &  -0.429 \\ 
H/Fe &  50 &  8 & -1.390 &   -0.299   & -0.200 &  -0.466 \\
H/Fe &  50 &  9 & -1.431 &   -0.255   & -0.217 &  -0.451 \\
H/Fe &  50 & 10 & -1.440 &   -0.235   & -0.210 &  -0.415 \\[5pt]
H/Fe &  70 &  7 & -1.320 &   -0.381   & -0.170 &  -0.413 \\
H/Fe &  70 &  8 & -1.311 &   -0.390   & -0.182 &  -0.425 \\
H/Fe &  70 &  9 & -1.305 &   -0.374   & -0.190 &  -0.445 \\
H/Fe &  70 & 10 & -1.315 &   -0.338   & -0.197 &  -0.461 \\[5pt]
H/Fe &  100&  7 & -1.315 &   -0.364   & -0.152 &  -0.376 \\
H/Fe &  100&  8 & -1.326 &   -0.392   & -0.163 &  -0.382 \\
H/Fe &  100&  9 & -1.323 &   -0.408   & -0.180 &  -0.428 \\
H/Fe &  100& 10 & -1.318 &   -0.419   & -0.191 &  -0.452 \\
\hline
\end{tabular}
\end{table}

We have also computed model atmospheres of a more complex structure,
in which pure iron gas is covered by a pure hydrogen layer of an arbitrary
thickness. Table 4 presents color indices of such stratified atmospheres,
in which we have arbitrarily assumed that the transition depth between
hydrogen and iron layers is located at the standard optical depth 
$\tau_0 = 10^{-3}$. This assumption precisely determines the column mass
of a hydrogen layer in each model atmosphere 
\begin{equation}
{\rm col. mass} = \int\limits_0^{\tau_0}
   {{d\tau_{std}} \over {(\kappa_\nu +\sigma_\nu)_{std}} } \,\,\, 
\label{eq:colmass}
\end{equation}
Hydrogen column mass in our H-Fe stratified atmospheres weakly
depends both on $\te$ and $\log g$ of the models. For example, H column 
mass is equal to $9.6025\times 10^{-5}$ g cm$^{-2}$ in the coldest model of
$\te = 20000$ K and $\log g = 8$, and slowly rises to $3.0602\times 10^{-4}$
g cm$^{-2}$ for the hottest model of $\te = 100000$ K and $\log g = 9$
(cf. sequence of sample models, Figs. 2-7). 

It would be very interesting to convert H column masses to the total
mass of a hydrogen envelope covering the iron core of a white dwarf. Such
a rescaling requires the knowledge of the radius $R$, whereas our models
are parametrized by the gravity, $g \sim M/R^2$. We have arbitrarily assumed
the mass of a white dwarf to be $M=M_\odot$, and used $M-R$ relations for
iron core
white dwarfs (Panei et al. 2000; Althaus 2003).
Respectively, the total mass of a hydrogen envelope in two above sample
models is lesser than $10^{-19} M_\odot$ for $\te = 20000$ K and 
$\log g = 8$, and is approximately equal to $2.6 \times 10^{-19} M_\odot$
for the hottest model of $\te = 100000$ K and $\log g = 9$.

One can note that the estimated masses of a hydrogen layer are rather
small, as compared with H masses arbitrarily assumed in published grids 
of theoretical $M-R$ relations ($M = 10^{-16} $ up to $10^{-4} M_\odot)$.
This is because H-Fe stratified models in the present paper require 
a H layer of the moderate optical depth $\tau_\nu < 1$ in frequencies 
corresponding to the UBVRI filters, which do not cover completely the
iron core.

Figs. 8 and 10 show positions of various model atmospheres on the theoretical 
$B-V$ vs. $U-B$ diagram, whereas Figs. 9 and 11 display colors of the
models on the $B-V$ vs. $V-I$ diagram. The Figures present colors of
both pure and stratified atmospheres. 
For these models, we obtained $U-B$, $B-V$ and $V-I$ colors which depend on
changes of $\te$ and $\log g$ in a rather complex way (cf. Tables 3--4).
One can clearly see that the colors of iron-containing
models may overlap, causing them to be a non-unique function
of the set of parameters $\te$, $\log g$. Pure hydrogen models are also
subject to this feature (at least on the $B-V$ vs. $V-I$ diagram),
but the complexity for the iron models is much higher. This complicate dependence
of the colors is due to a noisy structure of the TOPS grids of iron monochromatic opacities. 
Consequently, all these Figures present shaded areas, where atmospheres
containing pure iron are located. We do not highlight lines of constant 
$\te$ or $\log g$ for those models. One can see, that the region occupied
by pure iron model atmospheres significantly differs from the region 
corresponding to pure hydrogen models in all the analysed cases.

The latter observation shows that it would be possible to distinguish pure
iron WD atmospheres from pure hydrogen DA atmospheres with $\te \ge 20000$
K by means of the broadband Johnson UBV photometry, at least in principle.
Our analysis and Figures include neither possible admixture of 
helium or other heavier elements in DA white dwarf atmospheres, nor
the influence of their opacities on spectra and theoretical colors of mostly
hydrogen white dwarf atmospheres. Admixture of heavier elements in hot
WD atmospheres seems very common in real stars. For example, Barstow et al.
(1993) showed that most of the hot DA white dwarfs with effective temperatures
$\te > 40000 $ K have some abundance of elements heavier than He in their 
atmospheres. This can explain their generally low X-ray luminosity, as
it is found in the $ROSAT$ all-sky catalog of white dwarfs.

There exist many $U-B$ and $B-V$ color measurements available for isolated
white dwarfs, which were published in the catalogue by McCook \& Sion (1999).
Therefore we have plotted colors for individual white dwarfs in both
Figs. 8 and 10 as asterisks, in order to search for possible white
dwarfs with pure or stratified Fe atmospheres. Unfortunately, one
can note that the observed stars generally do not fall into shaded
areas of each Figure, but rather avoid them. Most of the observed
$U-B$ and $B-V$ colors mostly overlap the area of pure hydrogen atmospheres.

The above observation proves that the set of white dwarfs with color
indices predicted for iron atmospheres is almost empty, except for a few
cases of marginal significance.

Figs. 9 and 11 do not include observed colors of real WDs. Measurements
of near infrared luminosity $I$ (and therefore $V-I$ color) are extremely 
rare for hotter isolated white dwarfs. Also the newest catalogs of 
$B-V$ and $V-I$ of hot objects, which are known to us (Renzini 2001; 
Zoccali et al. 2001) do not distinguish hydrogen DA and other types of 
white dwarfs. Existing surveys of $V-I$ colors of white dwarfs
concentrate on the search of infrared excess presumably in binary stars,
and are not useful in this research.

   \begin{figure}[t]
    \resizebox{\hsize}{!}{\rotatebox{90}{\includegraphics{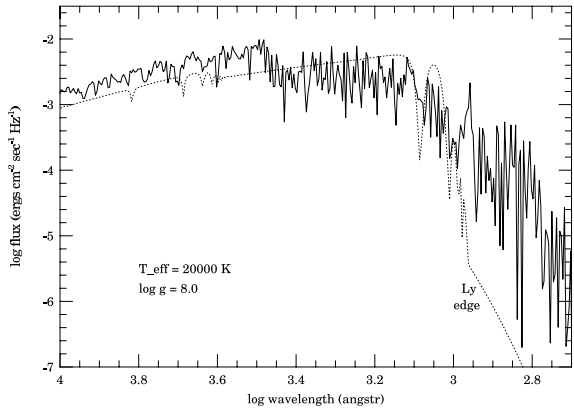}}}
    \caption[]{Pure iron and hydrogen LTE white dwarf synthetic spectra
     of $\te = 20000$ K and $\log g = 8.0$, displayed on wide range of 
     wavelengths from near infrared, $\lambda = 10000 $ {\AA }, down to about
     500 {\AA }. Solid line represents the spectrum of pure iron model and
     dotted line is the spectrum of pure H atmosphere. Both model spectra are
     computed using the EOS of nonideal gas and the Los Alamos TOPS 
     monochromatic opacities.  }
    \label{fig:fig2}
   \end{figure}

   \begin{figure}
    \resizebox{\hsize}{!}{\rotatebox{90}{\includegraphics{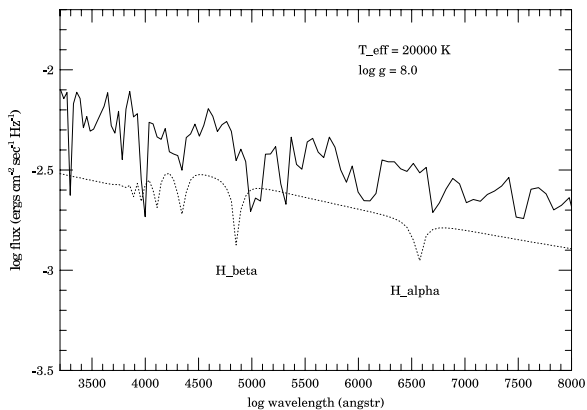}}}
    \caption[]{The LTE theoretical spectra of iron and hydrogen 
     atmospheres of Fig. 1 are displayed here on narrower range of wavelengths, with higher
     resolution, from near $\lambda = 3200 $ {\AA } up to 8000 {\AA }. 
     Again, solid line represents the spectrum of a pure iron model and 
     dotted line is the spectrum of a pure H atmosphere.  }
    \label{fig:fig3}
   \end{figure}

   \begin{figure}
    \resizebox{\hsize}{!}{\rotatebox{90}{\includegraphics{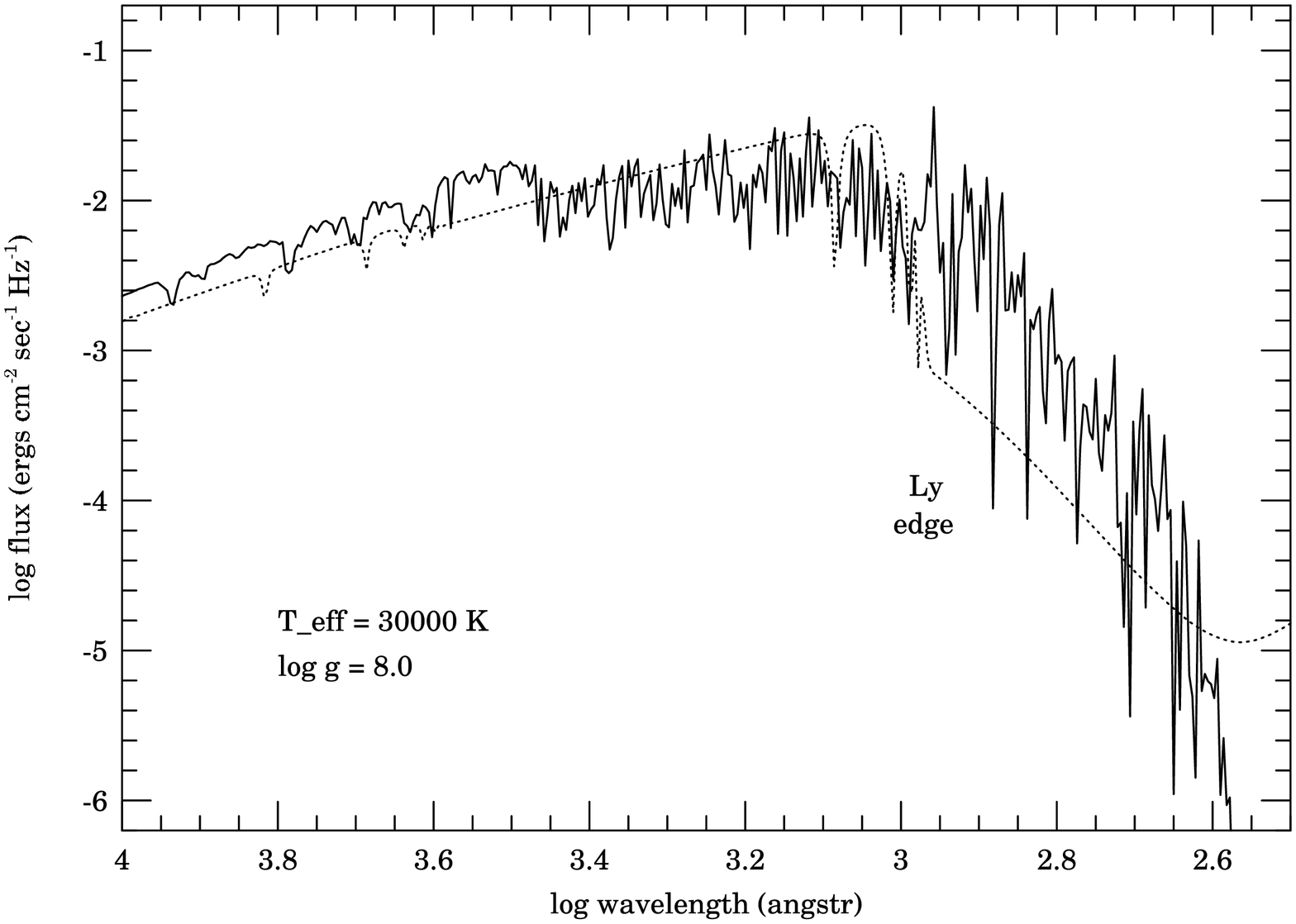}}}
    \caption[]{ Same as Fig. 2 for $\te = 30000$ K and $\log g = 8.0$.  }
    \label{fig:fig4}
   \end{figure}

   \begin{figure}
    \resizebox{\hsize}{!}{\rotatebox{90}{\includegraphics{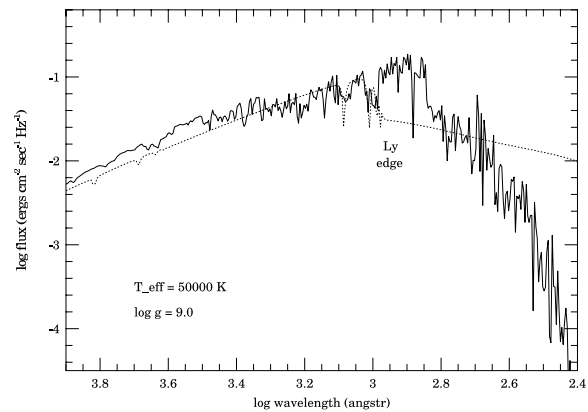}}}
    \caption[]{ Same as Fig. 2 for $\te = 50000$ K and $\log g = 9.0$. Note, that
     the surface gravity is higher than in previous Figures.  }
    \label{fig:fig5}
   \end{figure}

   \begin{figure}
    \resizebox{\hsize}{!}{\rotatebox{90}{\includegraphics{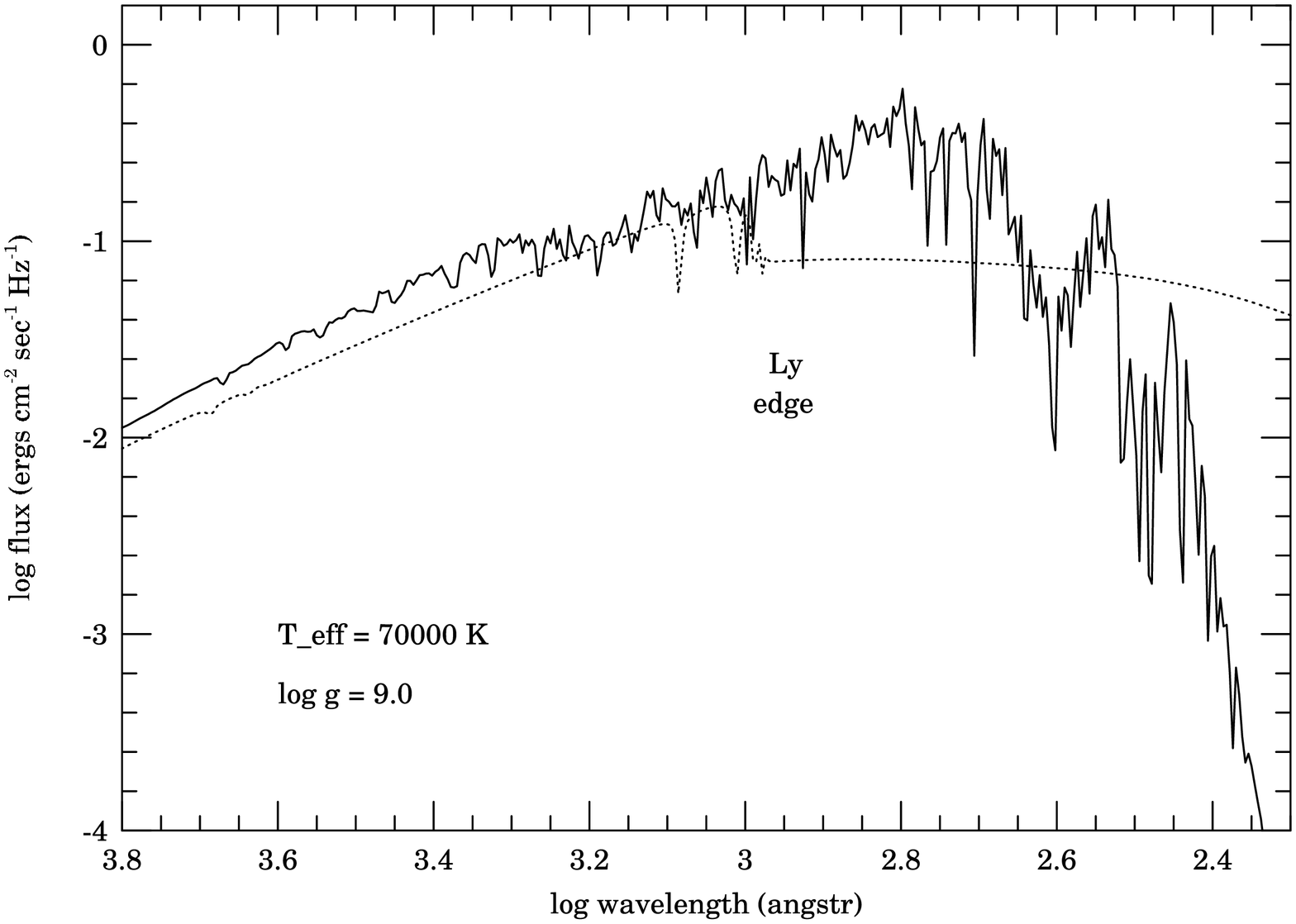}}}
    \caption[]{ Same as Fig. 2 for $\te = 70000$ K and $\log g = 9.0$.  }
    \label{fig:fig6}
   \end{figure}

   \begin{figure}
    \resizebox{\hsize}{!}{\rotatebox{90}{\includegraphics{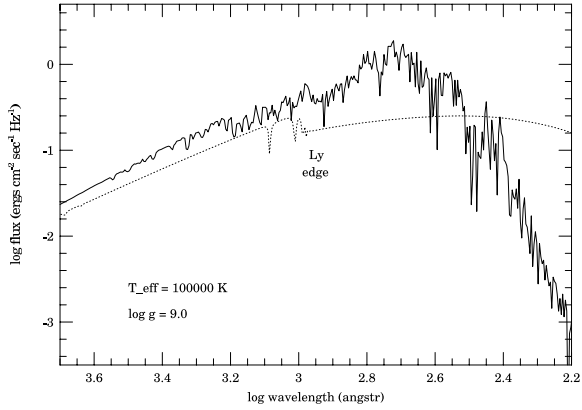}}}
    \caption[]{ Same as Fig. 2 for $\te = 100000$ K and $\log g = 9.0$.  }
    \label{fig:fig7}
   \end{figure}

\section{Search for iron core white dwarf candidates}

\subsection{Listing by Provencal et al. (1998) }

It is interesting to search for possible iron core white dwarf stars
among the existing WD catalogs. The first such effort has been done by 
Provencal et al. (1998), who compared masses and radii of
white dwarf stars with theoretical mass-radius relations for various 
WD core compositions. Their mass and radius determinations were
based mostly on the {\it Hipparcos} parallaxes, thus suggesting higher
accuracy. Mentioned paper includes also Procyon B as a candidate for
iron core star, however, this star has been classified as a normal white 
dwarf based on precise Hubble spectral measurements (Provencal et
al. 2002).

Based on the first quoted paper we have selected 7 white dwarfs, for which 
best determined masses and radii place them closest to the theoretical
mass-radius relation for iron core white dwarf stars (cf. Fig. 3 in
Provencal et al. 1998). Neglecting possible errors of $M$ and $R$
determinations we suggest that these stars are promising candidates
for comparing them with our theoretical UBVRI color indices (see Table 5). 
Some of these stars were also indicated by Panei et al. (2000), who 
placed them on their theoretical mass-radius relation for iron core
white dwarfs. 

Unfortunately, stars listed in our Table 6 are rather cold. The hottest of
them, EG 50 ($\te = 21000$ K) and GD 140 ($\te=21700$ K), are located
practically at the border of the grid presented in this paper. 
Color indices and the position of both stars in our two-color diagrams
places them far from the region occupied by pure iron atmospheres.
Moreover, they are located in the region occupied by 
pure hydrogen atmospheres. Therefore we conclude that even if those 
stars have iron cores, they are covered by a thick hydrogen layer of a 
large optical depth. This conclusion is supported by the fact that 
McCook \& Sion (1999) classified all the stars from Table 5 as pure hydrogen
DA white dwarfs.

\begin{table}
\caption{List of iron core candidate stars selected from the data by 
Provencal et al. (1998). Procyon B has been excluded already from their data.}
\label{tab:star}
\renewcommand{\arraystretch}{1.0}
\begin{tabular}{clllcr}
    \hline
WD &  $M/M_{\odot}$ & $\log g$ & $R/R_{\odot}$ & $U-B$ & $B-V$ \\
\hline \hline 
EG 21       & 0.554 & 8.06 & 0.0115  &  -0.55 &  0.05  \\
EG 50       & 0.497 & 8.10 & 0.0104  &  -0.91 & -0.08  \\
G 238-44    & 0.417 & 7.90 & 0.0120  &  -0.86 & -0.09  \\
G 181-B55   & 0.50  & 8.05 & 0.0110  &  -0.55 &  0.17  \\
GD 140      & 0.797 & 8.48 & 0.0085  &  -0.95 & -0.06  \\
GD 279      & 0.411 & 7.83 & 0.0129  &  -0.60 &  0.10  \\ 
WD 2007-303 & 0.433 & 7.86 & 0.0128  &  -0.66 &  0.07  \\
\hline
\end{tabular}
\end{table}

\subsection{ Other catalogs of WD masses and surface gravities }

In order to search for candidate iron core stars we have performed search of
papers listing mass and surface gravity determinations of white dwarfs 
carried out in the recent surveys (Vennes et al. 1997; Finley et al. 1997;
Homeier et al. 1998; Napiwotzki, Green \& Saffer 1999; Vennes 1999; 
Bergeron et al. 2001). 

The above papers discuss mostly DA white dwarfs,
i.e. those with hydrogen-rich atmospheres and therefore strong Balmer
lines (cf. Figs. 2--3). Grids of theoretical $UBVRI$ colors of Fe-abundant
atmospheres given in Tables 3--4 are not useful here. Therefore we attempted 
to identify stars of exceptionally high $\log g$ (and hence a small
radius) in those papers. Unfortunately, several trial figures displaying
$M$ vs. $\log g$ relations have shown that there are no stars which 
exhibit large $\log g$ for a given mass as compared with the general trend. 

As an example, we plot two such relations in Fig. 12, which are based on
the data from Finley et al. (1997), and Homeier et al. (1998). The former
paper presents a table with mass and surface gravity determinations for 90
hot DA white dwarfs from the {\sl Extreme Ultraviolet Explorer} ($EUVE$)
all-sky survey. 

\clearpage

  \begin{figure*}
    \resizebox{11cm}{!}{\includegraphics{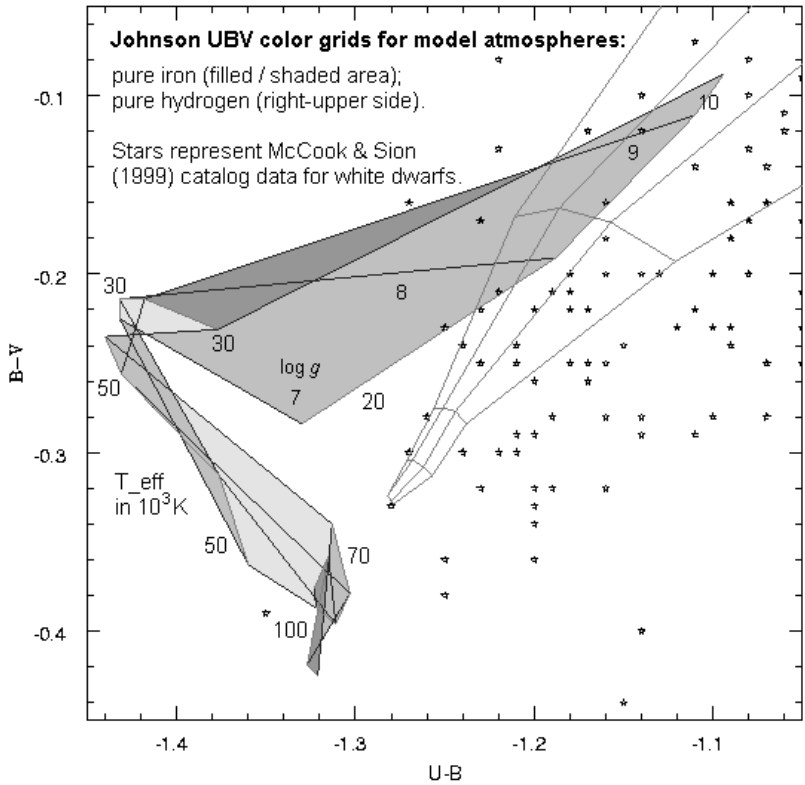}}
    \hfill
    \parbox[b]{55mm}{   
    \caption{ Plot of $U-B$ vs. $B-V$ for hot white dwarf stars.
              Shaded area denotes region occupied by colors of pure iron
              model atmospheres computed in this paper.  Numbers put in
              this area denote effective temperatures $\te$ (in
              thousands K) and $\log g$, respectively.  Colors of pure 
              hydrogen models (nonideal EOS) are represented by the
              transparent grid of sections.  Asterisks denote colors of
              real white dwarfs, as taken from McCook \& Sion (1999).
              One can see that only few real white dwarfs fall into the
              shaded area.
    }} 
   \end{figure*}

   \begin{figure*}
    \resizebox{11cm}{!}{\includegraphics{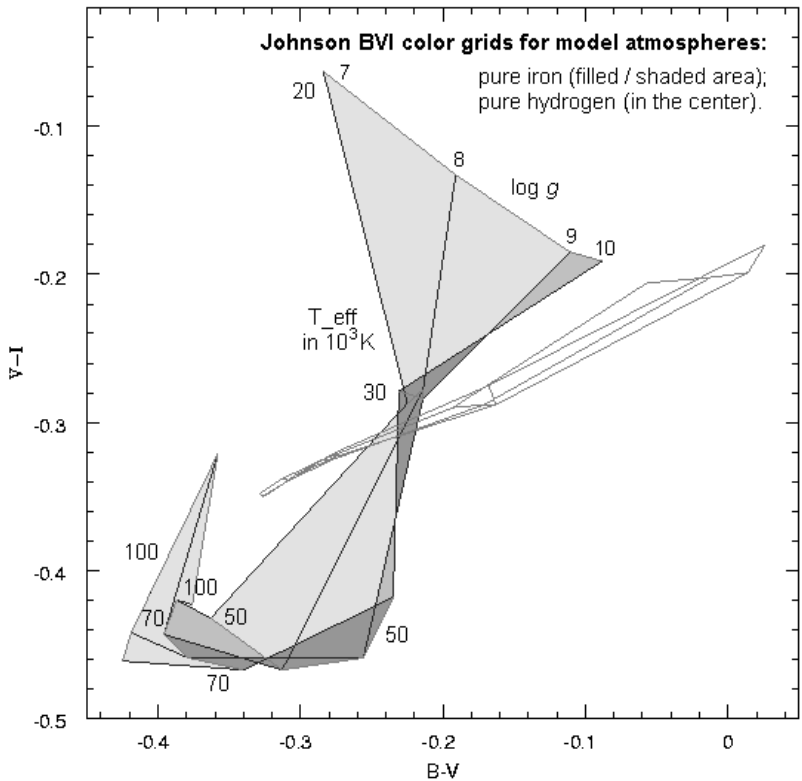}}
    \hfill
    \parbox[b]{55mm}{
    \caption{ Plot of $B-V$ vs. $V-I$ for hot white dwarf stars.  Also
              here, shaded area and the adjacent numbers correspond to
              the colors of pure Fe model atmospheres.  Transparent grid
              of sections correspond to pure H models (nonideal EOS).
              Unfortunately, there are no observed color indices available.    
}} 
   \end{figure*}

  \begin{figure*}
    \resizebox{11cm}{!}{\includegraphics{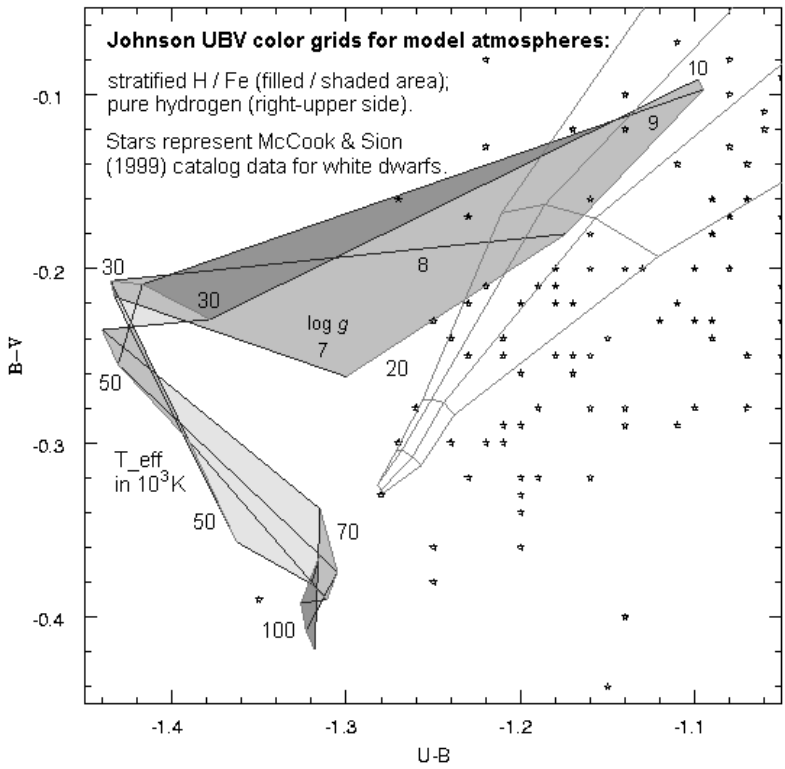}}
    \hfill
    \parbox[b]{55mm}{   
    \caption{ Plot of $U-B$ vs. $B-V$ for hot stratified Fe/H model
              atmospheres.  In these models, pure Fe atmosphere is
              covered by a pure H layer, reaching from the standard
              optical depth $\tau = 10^{-10}$ down to $10^{-3}$ in all
              cases.  Designations of $\te$ and $\log g$ have the same
              meaning as in previous Figures.  Pure H models are also
              represented in the same way.
}} 
   \end{figure*}

   \begin{figure*}
    \resizebox{11cm}{!}{\includegraphics{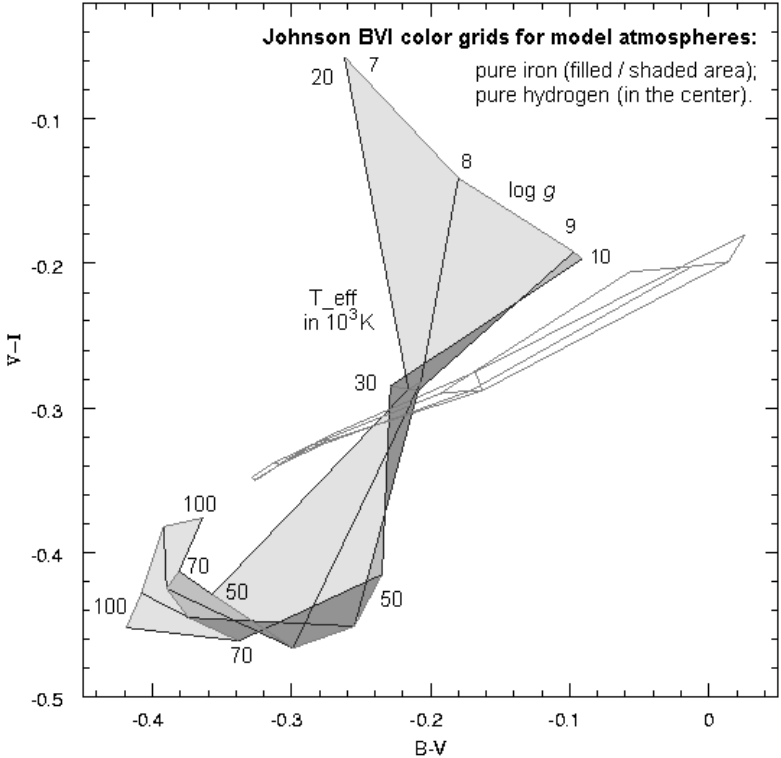}}
    \hfill
    \parbox[b]{55mm}{
    \caption{ Plot of $B-V$ vs. $V-I$ for Fe/H stratified atmospheres.
              Again, no observations of real white dwarfs are available 
              here.     
   }} 
   \end{figure*}

\clearpage

Surface gravity $\log g$ and $\te$ are usually determined by fitting of 
the observed Balmer line profiles and therefore this method makes direct
to observable quantities. One should necessarily note, however, that in 
all the above papers masses of DA white dwarfs were obtained using the
theoretical mass-radius relations for carbon-core stars with hydrogen or
helium envelope (see Wood 1995). We are aware that if masses were
determined with the most new mass-radius relation for iron-core WDs by 
Panei et al. (2000), then the above Figure would change significantly.

\begin{figure}
  \resizebox{\hsize}{!}{\rotatebox{90}{\includegraphics{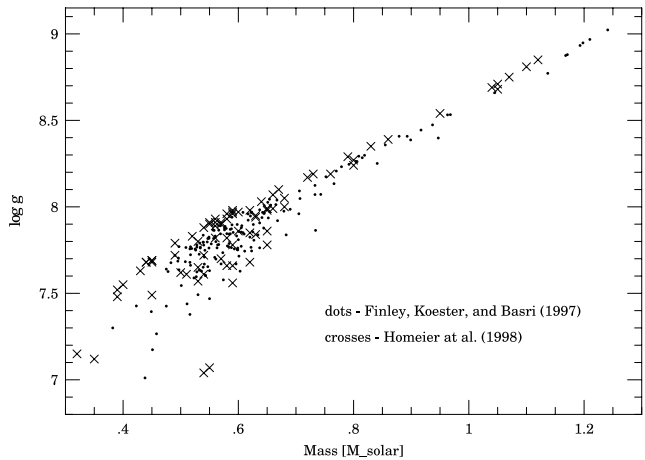}}}
  \caption[]{Catalog of $M/M_{\odot}$ vs. $\log g$ determinations for two
  sets of DA white dwarfs. Dots denote stars observed by $EUVE$ satellite
  by Finley, Koester, \& Basri (1997). Crosses represent data from the
  Hamburg Quasar Survey (Homeier et al. 1998).
  All the investigated stars are located in well
  defined strip corresponding to C-O white dwarf cores. No single star with
  too high $\log g$ is seen in the Figure.}
 \label{equ:fig4}
\end{figure}

\section{Summary}

The most important areas of concern of our paper can be summarized as follows.

{\bf 1.} 
We have computed and presented here LTE model atmosphere computations 
of hot white dwarfs, which atmospheres consist either of pure iron,
stratified iron/hydrogen layers, or pure hydrogen. Model atmospheres are
in both radiative and hydrostatic equilibrium. Computations use exclusively
the equation of state for nonideal Fe and H gases, obtained by the OPAL project,
and the extensive tables of the TOPS monochromatic opacities from the Los
Alamos National Laboratory. The latter opacities are fully compatible with
the nonideal EOS for both gases.

{\bf 2.} 
This paper presents calibration of colors $U-B$, $B-V$, $V-R$ and $V-I$
of the broadband Johnson $UBVRI$ photometry, for pure Fe, pure H and
stratified Fe/H atmospheres. These colors were computed
from theoretical spectra of our model atmospheres with a nonideal EOS.

{\bf 3.}
The paper investigates the influence of nonideal equation of state on the
photometric $UBVRI$ colors of the broadband Johnson system, both in pure
H and pure Fe model atmospheres. We showed that effects of nonideal EOS
for pure iron model atmospheres are substantial in the investigated
models, contrary to pure H WD atmospheres.   

{\bf 4.}
We have also computed $UBVRI$ colors of the models and
determined the area on the $B-V$ vs. $U-B$ and $B-V$ vs. $V-I$ planes, occupied
by both pure Fe and pure H model atmospheres of WD stars. 

\vb
The importance of our paper is that we have determined the area on
two-color diagrams in which colors of the hypothetical pure iron
or stratified H/Fe atmospheres should be located. Colors of the above
hypothetical atmospheres are distinctly different from colors of pure
hydrogen DA WSs with $\te \ge 20000$ K. We are aware, however, that
our paper does not predict colors of iron core WDs in case, if the
core is covered by a hydrogen layer of significant optical thickness.

\begin{acknowledgements}

We thank I. Hubeny for his agreement to use the {\sc tlusty195} code and
related discussions.
We appreciate also comments by P. Bergeron and A. Renzini regarding photometric
properties of DA white dwarfs. 
Our thanks are also due to L.G. Althaus for providing us with his
theoretical $M - R$ relations for various white dwarfs.
This work has been supported by the Polish Committee for
Scientific Research grant No. 2 P03D 021 22.

\end{acknowledgements}


\begin{thebibliography}{}

\bibitem[]{} Althaus, L.G., 2003, personal communication

\bibitem[]{} Barstow, M.A., Fleming, T.A., Diamond, C.J., Finley, D.S.,
    Sansom, A.E., Rosen, S.R., Koester, D., Holberg, M.C., Marsh, J.B.,
    Kidder, K., 1993, MNRAS, 264, 16

\bibitem[]{} Bergeron, P., Wesemael, F., Beauchamp, A., 1995, PASP,
    107, 1047

\bibitem[]{} Bergeron, P., Leggett, S.K., Ruiz, M.T., 2001, ApJS, 133, 413

\bibitem[]{} Bessell, M.S., 1990, PASP, 102, 1181

\bibitem[]{} Castelli, F., Kurucz, R.L., 1994, A\&A, 281, 817

\bibitem[]{} Finley, D.S., Koester, D., Basri, G., 1997, ApJ, 488, 375

\bibitem[]{}
Girard, T.M., Wu, H., Lee, J.T., Dyson, S.E., van Altena, W.F.,
Horch, E.P., Gilliland, R.L., Schaefer, K.G., Bond, H.E., Ftaclas, C., 
Brown, R.H., Toomey, D.W., Shipman, H.L., Provencal, J.L., Pourbaix, D.,
    2000, AJ, 119, 2428

\bibitem[]{}
Girardi, L., Bertelli, G., Bressan, A., Chiosi, C., Groenewegen, M.A.T.,
    Marigo, P., Salasnich, B., Weiss, A., 2002, A\&A, 391, 195

\bibitem[]{} Hayes, D.S., 1985, Calibration of fundamental stellar
    quantities, IAU Symposium 111, eds. D.S. Hayes, L.E. Pasinetti,
    and A.G.D. Philip (Dordrecht, Reidel), 225

\bibitem[]{} Homeier, D., Koester, D., Hagen, H.-J., Jordan S., Heber, U.,
    Engels, D., Reimers, D., Dreizler, S., 1998, A\&A 338, 563

\bibitem[]{} Hubeny, I., 1988, Comput. Phys. Comm., 52, 103

\bibitem[]{} Hubeny, I., Lanz, T., 1992, A\&A, 262, 501

\bibitem[]{} Hubeny, I., Hummer, D.G., Lanz, T., 1994, A\&A, 282, 151

\bibitem[]{} Hubeny, I., Lanz, T., 1995, ApJ, 439, 875

\bibitem[]{} Lanz, T., Hubeny, I., 1995, ApJ, 439, 905

\bibitem[]{} Lemke, M., 1997, A\&AS, 122, 285

\bibitem[]{} Madej, J., 1991, ApJ, 376, 161

\bibitem[]{} Madej, J., 1994, Acta Astron. 44, 191

\bibitem[]{} Madej, J., 1998, A\&A, 340, 617

\bibitem[]{} Madej, J., R\'o\.za\'nska, A., 2000, A\&A, 356, 654 
    
\bibitem[]{} Magee, N.H., Abdallah, J., Jr., Clark, R.E.H.,
   Cohen, J.S., Collins, L.A., Csanak, G., Fontes, C.J., Gauger, A., 
   Keady, J.J., Kilcrease, D.P., Merts, A.L., 1995,
   Astrophysical Applications of Powerful New Databases. 
   Joint Discussion No. 16 of the 22nd. General Assembly
   of the I.A.U., ASP Conference Ser., Vol. 78, 51
   eds. S.J. Adelman \& W.L. Wiese

\bibitem[]{} McCook, G.P., Sion, E.M., 1999, ApJS, 121, 1

\bibitem[]{} Mihalas, D. 1978, Stellar Atmospheres, 2nd Ed., W.H. Freeman \&
     Co., San Francisco

\bibitem[]{} Napiwotzki, R., Green, P.J., Saffer, R.A., 1999, ApJ, 517, 399 

\bibitem[]{} Panei, J.A., Althaus, L.G., Benvenuto O.G., 2000, A\&A,
   353, 970

\bibitem[]{} Provencal, J.L., Shipman, H.L., 1999, 11th European
   Workshop on White Dwarfs, ASP Conference Series, Vol. 169, 293, eds. 
   J.-E. Solheim \& E.G. Meistas 

\bibitem[]{} Provencal, J.L., Shipman, H.L., Hog, E., Thejll, P.,  
   1998, ApJ, 494, 759

\bibitem[]{} Provencal, J.L., Shipman, H.L., Koester, D., Wesemael, F., 
   Bergeron, P. 2002, ApJ, 568, 324

\bibitem[]{} Renzini, A., 2001, personal communication

\bibitem[]{} Rogers, F.J., 1994, in ``The Equation of State in Astropyhsics'',
   IAU Colloquium 147, eds. G. Chabrier and E. Schatzman (Cambridge University
   Press), p. 16

\bibitem[]{} Rogers, F.J., 2002, personal communication

\bibitem[]{} Rogers, F.J., Swenson, F.J., Iglesias, C.A., 1996,
   ApJ, 456, 902

\bibitem[]{} Rohrmann, R.D., Serenelli, A.M., Althaus, L.G., Benvenuto, O.G.,
   2002, MNRAS, 335, 499

\bibitem[]{} Stage, M.D., Joss, P.C, Madej, J., 2002, Neutron Stars
   in Supernova Remnants, ASP Conference Series, Vol. 271, 327,
   eds. Patrick O. Slane \& Bryan M. Gaensler

\bibitem[]{} Vennes, S., 1999, ApJ, 525, 995

\bibitem[]{} Vennes, S., Thejll, P.A., Galvan, R.G., Dupuis, J.,
   1997, ApJ, 480, 714

\bibitem[]{} Wesemael, F., Auer, L.H., Van Horn, H.M., Savedoff, M.P.,
   1980, ApJS, 43, 159

\bibitem[]{} Wood, M., 1995, Lecture Notes in Physics, 443, White Dwarfs,
   eds. D. Koester \& K. Werner (Berlin: Springer), 41

\bibitem[]{} Zoccali, M., Renzini, A., Ortolani, S., Bragaglia, A., 
   Bohlin, R., Carretta, E., Ferraro, F.R., Gilmozzi, R., Holberg, J.B.,
   Marconi, G., Rich, R.M., Wesemael, F., 2001, ApJ, 553, 733


\end{thebibliography}
\end{document}